\documentclass[10pt,conference]{IEEEtran}

\IEEEoverridecommandlockouts
\usepackage{cite}
\usepackage{amsmath,amssymb,amsfonts}
\usepackage{algorithmic}
\usepackage{graphicx}
\usepackage{textcomp}
\usepackage{xcolor}
\def\BibTeX{{\rm B\kern-.05em{\sc i\kern-.025em b}\kern-.08em
    T\kern-.1667em\lower.7ex\hbox{E}\kern-.125emX}}

\usepackage{xcolor}

\usepackage{tabularx}
\usepackage{colortbl}

\usepackage{listings}
\usepackage{url}
\usepackage{hyperref}

\usepackage{graphicx}
\usepackage{array}			
\usepackage{multirow}
\usepackage{tabularx}
\usepackage{booktabs} 

\usepackage{enumitem}

\usepackage{tikz}

\usepackage{ifthen}
\usepackage{amssymb}
\newboolean{showcomments}
\setboolean{showcomments}{true} 
\ifthenelse{\boolean{showcomments}}
{\newcommand{\nb}[2]{
    \fcolorbox{gray}{yellow}{\bfseries\sffamily\scriptsize#1}
    {$\blacktriangleright$#2$\blacktriangleleft$}
  }
  
}
{\newcommand{\nb}[2]{}
  
}

\newcommand{\eg}{e.g.,~}							
\newcommand{\ie}{i.e.,~}							
\newcommand{\Fig}[1]{Figure~\ref{#1}}  			
\newcommand{\Table}[1]{Table~\ref{#1}}	    
\newcommand{\Sect}[1]{Section~\ref{#1}}	  
\newcommand{\oapp}{DISCOREV}
\newcommand*\circled[1]{\tikz[baseline=(char.base)]{
            \node[shape=circle,text=black!100, draw,inner sep=1pt] (char) {#1};}}

\begin{document}

\title{Unity is Strength: Cross-Task Knowledge Distillation to Improve Code Review Generation
\thanks{The Replicate package is made available in the supplementary material of this paper.}
}

\author{
\IEEEauthorblockN{Oussama Ben Sghaier}
\IEEEauthorblockA{\textit{University of Montreal}\\
Montreal, Canada \\
oussama.ben.sghaier@umontreal.ca}

\and
\IEEEauthorblockN{Lucas Maes}
\IEEEauthorblockA{\textit{University of Montreal}\\
Montreal, Canada \\
lucas.maes@umontreal.ca}

\and
\IEEEauthorblockN{Houari Sahraoui}
\IEEEauthorblockA{\textit{University of Montreal}\\
Montreal, Canada \\
houari.sahraoui@umontreal.ca}
}


\maketitle

\begin{abstract}

Code review is a fundamental process in software development that plays a critical role in ensuring code quality and reducing the likelihood of errors and bugs. However, code review might be complex, subjective, and time-consuming. Comment generation and code refinement are two key tasks of this process and their automation has traditionally been addressed separately in the literature using different approaches. In this paper, we propose a novel deep-learning architecture, \oapp, based on cross-task knowledge distillation that addresses these two tasks simultaneously. In our approach, the fine-tuning of the comment generation model is guided by the code refinement model. We implemented this guidance using two strategies, feedback-based learning objective and embedding alignment objective. We evaluated our approach based on cross-task knowledge distillation by comparing it to the state-of-the-art methods that are based on independent training and fine-tuning. Our results show that our approach generates better review comments as measured by the BLEU score.

\end{abstract}



\maketitle
\section{Introduction}

Code review is a fundamental process of the software development life cycle that aims to identify issues and sub-optimal code, detect bugs \cite{mcintosh2014impact, mcintosh2016empirical}, and ensure the quality of the source code \cite{ackerman1989software, ackerman1984software, morales2015code}. It is a manual process where one or more developers inspect the code written by fellow developers\cite{fagan2002design, bavota2015four}. Code review entails three main tasks: issue identification, issue description (i.e., review comment), and code refinement. 
Issue identification involves the localization of defects or problems within the source code, while the review description entails writing a comment that pinpoints the specific change that introduced the issue and may suggest a course of action for resolving it \cite{mcintosh2014impact, mcintosh2016empirical}. Code refinement denotes the correction stage in which the developer addresses the issue to fix it and fulfill the reviewer's comment\cite{bacchelli2013expectations}.

The code review process is often regarded as arduous, time-consuming, and intricate, particularly for large-scale projects\cite{eick2001does, avgeriou2016managing}. It is also a highly subjective process that can be influenced by various human and social factors (e.g., developer's experience, personal relationships, etc.). These subjective factors may lead to biased review comments being provided. These limitations may result in potential inefficiencies and inconsistencies in the code review process, potentially having a significant impact on the quality and reliability of the resulting codebase.

To address these challenges, there has been increasing interest in assistance and automation approaches to code review\cite{li2022codereviewer, tufano2022using, tufan2021towards, siow2020core, gupta2018intelligent, hovemeyer2004finding}.
One research area investigates the initial stages of the software development process by utilizing static analysis to identify potential issues \cite{wichmann1995industrial}. These tools, referred to as linters, define a set of manual rules representing different issues, and are used to flag parts of the code that contravene the predefined rules. However, the effectiveness of linters is limited, as the rules are manually defined and require substantial adaptation to cover various issues. Moreover, software issues are subject to changes over time and can be influenced by multiple factors such as software architecture, team, project, and domain. Consequently, this rigidity in static analysis tools curtails their widespread applicability and effectiveness in software projects \cite{bielik2017learning, sadowski2015tricorder}.
Other works employed similarity techniques \cite{siow2020core, gupta2018intelligent} to recommend relevant comments, from a pre-defined dataset of review comments, that correspond to changes in a code snippet. These approaches assume that reviewers may come across analogous situations to those they have previously encountered. Although such suggestions could prove useful, review comments are rarely generic, and more often relate to a specific context.

With the recent advancements in deep learning and natural language processing, there has been significant interest in using pre-trained language models to solve downstream tasks in software engineering. Accordingly, recent works \cite{tufano2022using, tufan2021towards, li2022codereviewer, sghaier2023multi} focus on using generative deep learning models to automate code review tasks including comment generation (\ie generate a review comment given the code changes) and code refinement (\ie fix the code to conform to the review comment). The proposed approaches improved dramatically the generation of code reviews. Notwithstanding the usefulness of these approaches and their promising results, code review tasks were conventionally considered separately and addressed independently despite their substantial interdependence.

In this paper, we rely on the strong relationship between the two tasks of code review, \ie comment generation and code refinement. These two tasks share knowledge since the output comment from the first task is fed to the code refinement task.
That is, code refinement consists of fixing the code to satisfy the review comment at hand while comment generation includes writing a description of the issue to drive the code refinement phase. 

Based on this statement, we propose a novel deep-learning architecture, called \oapp~(\emph{cross-task knowledge DIStillation for COde REView generation}), that incorporates two models trained jointly on the tasks of comment generation and code refinement. 
Our proposed architecture is based on cross-task knowledge distillation, call it \oapp, where one model gets the feedback of another to achieve better performance. 
The first model takes as input the source code and generates a review comment, which is fed along with the input source code to a second model to predict the refined version of the source code. 
This latter returns feedback that represents the relevancy and the informativeness degree of the generated comment.
By jointly training the two models, we can capture complex interactions, exchange knowledge and feedback, and ensure consistency between the code and the review comment.
Furthermore, we introduce an embedding alignment objective to enforce closer semantic representations for the comment and code.

\oapp~builds on the state of the art in deep learning for code review, which has recently shown promising results in various sub-tasks such as comment generation and code refinement \cite{li2022codereviewer, tufano2022using}. In addition to the reuse of pre-trained models, our approach ensures joint modeling of the comment generation and the code refinement tasks, which enables a more effective comment generation.

To evaluate \oapp, we conduct experiments using the same dataset of code reviews employed in the literature\cite{li2022codereviewer}. Our results show that our architecture outperforms state-of-the-art approaches for the comment generation task as measured by the BLEU score. 
Additionally, we investigate the impact of the embedding alignment objective on the model performance.
We found that enforcing close semantic representations between the comment and the code edits helps the model achieve better performance by generating more accurate and relevant comments.

The subsequent sections of this paper are structured as follows. 
\Sect{sec:background} provides a concise overview of the background.
\Sect{sec:approach} details the different components of our proposed approach.
\Sect{sec:evaluation} reports the evaluation results and discusses some threats related to our experiments.
\Sect{sec:literature} outlines some related work.
Finally, \Sect{sec:conclusion} concludes.

\section{Background}
\label{sec:background}

\subsection{Code review}
\label{Sec:code_review}
\begin{figure}[h]
  \centering
  \includegraphics[width=1\linewidth]{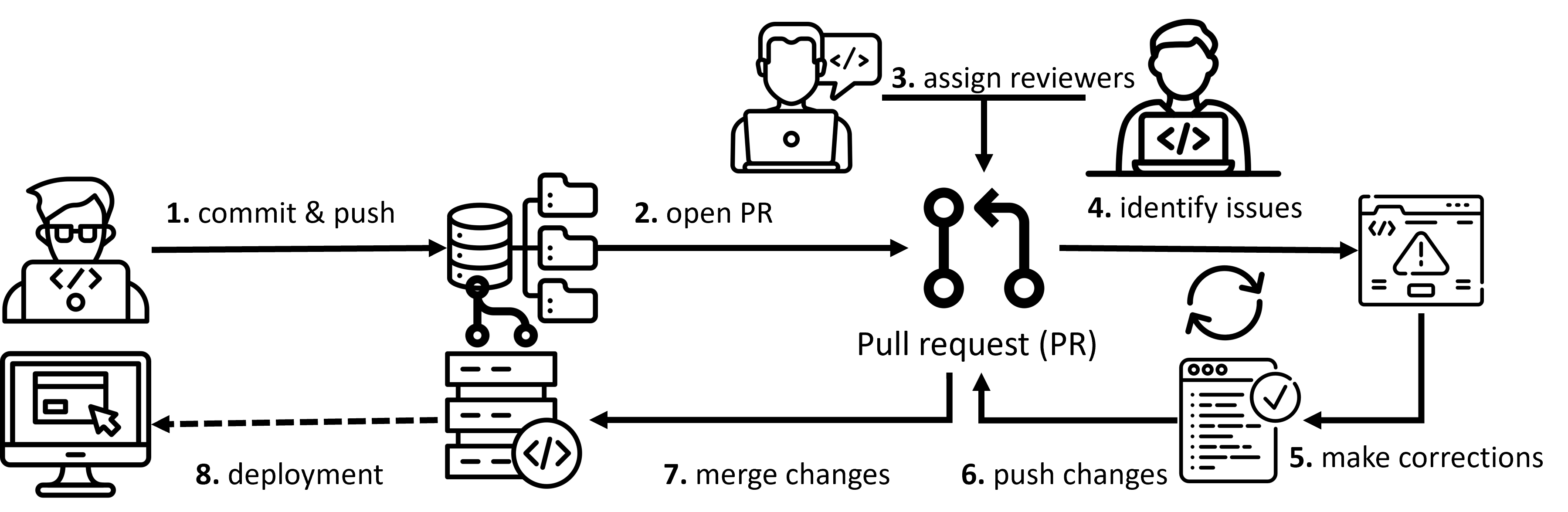}
  \caption{Conventional software development workflow in a continuous integration context}
  \label{fig:pr1}
\end{figure}

In a continuous integration setting, software developers utilize version control systems, such as GitHub, to collaborate and oversee software versions. 
As developers collaborate and persistently make changes to the codebase (\ie write new code or modify existing code), the use of version control systems enables development teams to efficiently manage and monitor the software codebase across time  \cite{shahin2017continuous, fowler2006continuous}. 
\Fig{fig:pr1} illustrates the conventional software development workflow in a continuous integration context.

A developer, who is working on implementing a new feature, makes local changes to the codebase.
Then, she commits and pushes her changes to a shared repository \circled{1} to make the new changes available to collaborators.
She can create a pull request \circled{2} to request that her changes be merged to the main branch (\ie main version of the source code). 
Some reviewers are assigned to the pull request \circled{3} to inspect the code changes made by the author before these changes are merged with the main branch. 
Reviewers identify and locate issues \circled{4} that need to be solved by the code author (\circled{5} and \circled{6}). 
The revised version of the code, submitted by developers, should be reviewed again.
Once these changes are approved by reviewers, they are merged  with the main code version, \ie main branch \circled{7}, to be available to other developers (\ie collaborators).
If necessary, the project can be deployed \circled{8}, so that changes will be available to end-users.

\begin{figure}[h]
  \centering
  \includegraphics[width=1\linewidth]{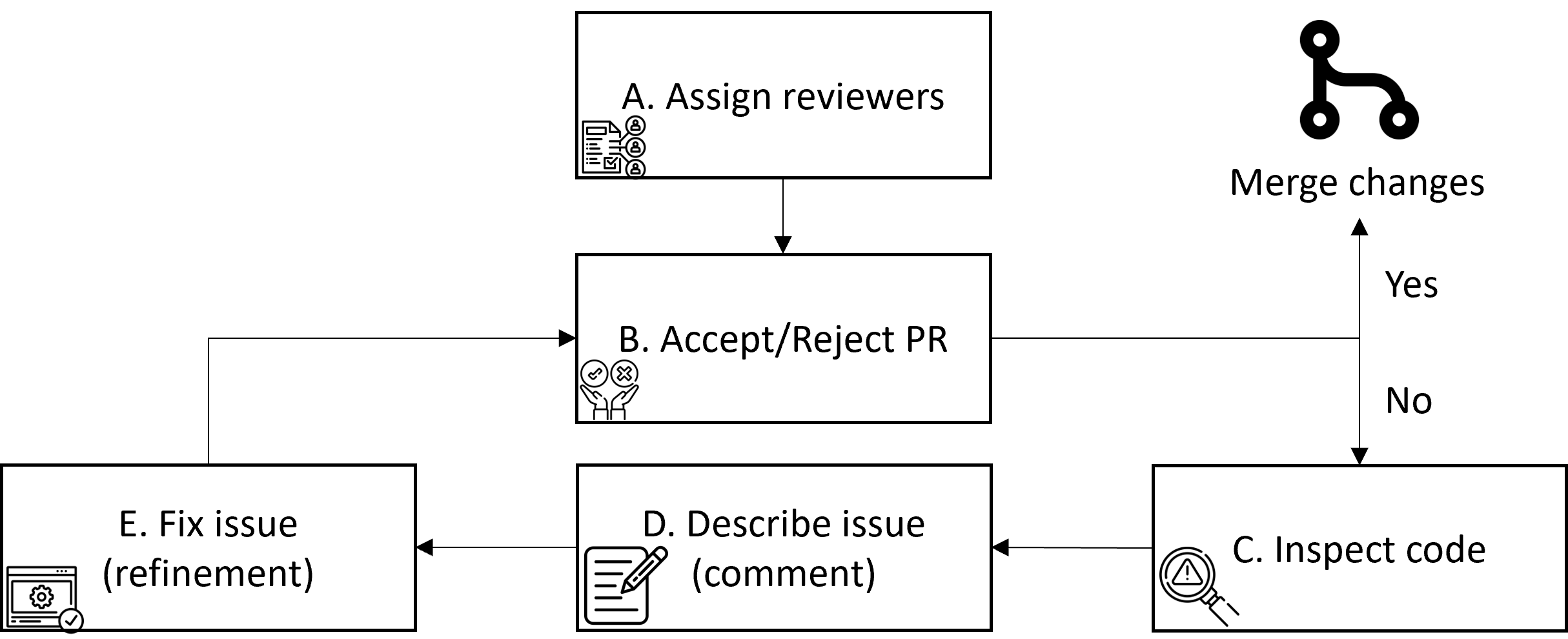}
  \caption{Process of code review}
  \label{fig:cdprocess}
\end{figure}

Code review is an essential task in the software development life cycle that aims to preserve the high quality of the software's codebase.
It entails regularly inspecting the source code written by fellow developers with the objective of identifying bugs, potential issues, sub-optimal code fragments, violations of code style, etc.
\Fig{fig:cdprocess} depicts an overview of this process.

The process of code review is triggered once a developer pushes her code to a remote shared repository.
The developer creates a pull request, \ie asks to merge her changes with the main codebase to make them available.
Prior to the merge with the main branch, these changes need to be inspected by going through a code review process.
The first step of this process involves assigning one or more reviewers to the pull request \circled{A}. This phase depends on many factors including expertise, availability, etc.
If the code is well-written and does not have problems \circled{B}, the pull request is accepted and the code changes are merged with the main branch of the codebase.
Conversely, if the code has issues, the reviewer should identify them \circled{C}.
She writes a comment to describe the issues at hand \circled{D} and potentially suggests some solutions.
Subsequently, the developer should address the comment by fixing the issues and applying the appropriate corrections\circled{E}. 
This is a recurring process where the developer should push the new changes to be checked again by the reviewers. 
If the issues are properly fixed, the reviewer accepts the pull request.
Otherwise, she could give further explanations or point out newly introduced issues.

\paragraph*{\textbf{Code review comments}}
Having assigned a reviewer to the pull request, her role is to inspect the newly introduced changes to the codebase. The reviewer examines the source code changes looking for issues or potential improvements. Issues might be related to security vulnerabilities, code style, code standards violations, inconsistencies, bugs, sub-optimal code fragments, etc. Reviewers can also suggest improvements that involve code efficiency and complexity, software libraries, best practices, refactoring, code quality, code optimization, documentation, etc. The identified issues and improvements are expressed by the mean of comments. The comments may be descriptive where the issues are detailed. However, comments may also be actionable where the reviewers recommend solutions to fix the issues. 
Code review comments are very useful as they enable developers, \ie authors of the code changes, to get feedback from experienced developers. This leads to better collaboration, communication, and knowledge-sharing within a team while maintaining a good codebase quality. This helps also avoid technical debt, which refers to the additional cost and effort required to fix issues that are not caught early in the development process.
Automating code review comments generation can bring several benefits to the software development process. 
Firstly, it can help save time and effort by reducing the amount of manual work needed to review and provide feedback on code. This enables teams to move faster in their development process.
The automation of this process helps to produce objective feedback without being influenced by personal preferences, relationships, health, well-being, etc.

\paragraph*{\textbf{Code refinement}}
The comments that are produced by the previous step are fed as input to the code refinement phase. The developers, and authors of the code changes, should consider the feedback of the reviewers and perform the necessary code changes. Developers should solve the issues described in the comments or apply suggested solutions. New code changes are pushed again to the project repository and should satisfy the reviewer's comments. The refined version of the source code is examined again by the reviewers to check whether the comments are properly addressed. Proper code refinement requires a good understanding of the reviewer's comments. That is, the reviews provided should be clear, specific, and comprehensible.

\subsection{Pre-trained language models}

In recent years, the Natural Language Processing (NLP) domain has dramatically evolved. 
State-of-the-art language models, that are based on the transformer architecture \cite{vaswani2017attention}, have shown outstanding performance in solving natural language problems.
Transformers have several advantages over recurrent neural networks. They enable parallel computations which reduces training time and captures long-range dependencies efficiently.
\emph{BERT}\cite{devlin2018bert}, \emph{GPT-3}\cite{radford2019language}, \emph{XLNet} \cite{yang2019xlnet}, \emph{RoBERTa} \cite{liu2019roberta} and \emph{T5} \cite{raffel2019exploring} are examples of general-purpose transformers that were pre-trained on tons of data to provide general and high-level representations of text.
\emph{CodeBERT} \cite{feng2020codebert} and \emph{CodeT5} \cite{wang2021codet5} are examples of transformers that were pre-trained on source code data.
These pre-trained models could then be adapted (\ie fine-tuned) on downstream tasks using specific datasets (\eg fine-tune a language model to predict if a code fragment contains bugs).

Transformers are sequence-to-sequence models based on the attention mechanism that takes into account the relationship between all the words in the sentence and not only the contextual words (\eg previous and next words).
The attention mechanism learns a weighting function indicating how much each element contributes to the prediction (\ie importance of each input element in predicting the target). 
A transformer is composed of encoders and decoders.
The encoder transforms the input sequence into contextual embeddings. 
These latter encode the meaning of each word of the input based on its context in low-dimension and high-level representations.
The decoder uses these contextual representations to predict the output sequence.

\emph{CodeT5} \cite{wang2021codet5} is a pre-trained encoder-decoder transformer for code that is pre-trained on $8.35M$ functions in 8 programming languages.
There are several versions of \emph{codeT5}. The small version has $60$ million parameters and the base version has $220$ million parameters.

\subsection{Knowledge distillation}
\label{Sec:distillation}
Knowledge distillation is a popular technique that was first introduced by Hinton et al. \cite{hinton2015distilling} as a method to transfer knowledge from a complex model, also known as the teacher, to a smaller and faster model, called the student. The main goal of knowledge distillation is to transfer the learned knowledge of the teacher model to the student model so that it can achieve comparable or even better performance on a target task.

The process of knowledge distillation consists of training a teacher model on a large dataset to accomplish a specific task. Then, the student model is trained to mimic the behavior of the teacher model by minimizing the distance between their output distributions on the same dataset, typically using the Kullback-Leibler divergence (KL-divergence) as the distance metric \cite{joyce2011kullback, kim2021comparing}.

For instance, this concept is employed in model compression where the knowledge is transferred from a large and complex neural network (\ie the teacher) to a smaller and simpler neural network (\ie the student). The goal of knowledge distillation is to make the student network learn the same function as the teacher network but with fewer parameters. 

Cross-task knowledge distillation is a recent extension of knowledge distillation that enables the transfer of knowledge from one task to another\cite{ye2020distilling, yuan2019ckd, yang2022cross, li2022prototype}. Instead of transferring knowledge from a complex model to a smaller one for the same task, cross-task knowledge distillation transfers knowledge from a model trained on one task, referred to as the source task, to a model trained on a different task, referred to as the target task. This allows the target model to improve its performance, even if the two tasks are not directly related.

\section{Proposed approach}
\label{sec:approach}

\subsection{Overview}
\begin{figure*}[!htbp]
    \centering
    \includegraphics[scale=0.1]{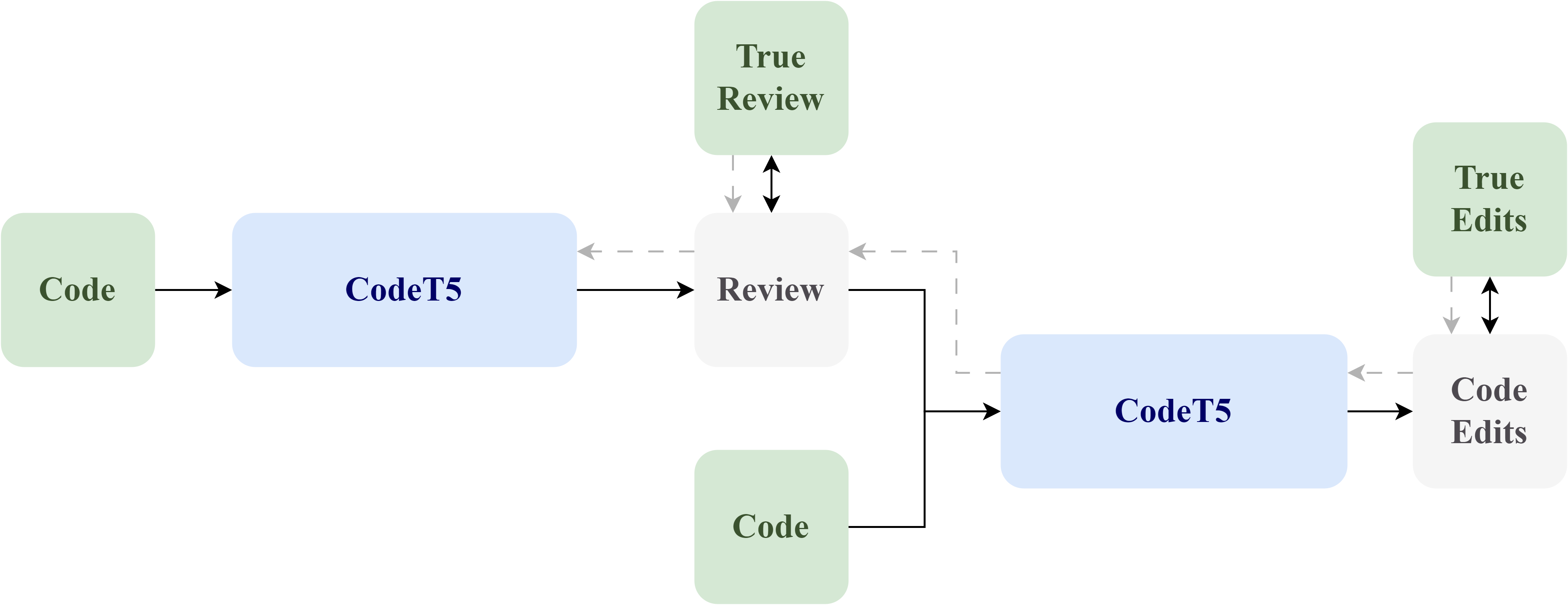}
    \caption{Overview of the proposed architecture \oapp. \\
    The forward arrows represent the forward pass, the backward dashed arrows represent the backpropagation of the model, and the two-sided arrows represent the loss function that compares the predictions to the ground truth}.
    \label{fig:architecture}
\end{figure*}

\Fig{fig:architecture} depicts an overview of the proposed approach.
We propose a novel deep learning architecture to assist an important task in the code review process, namely comment generation.
To efficiently address this, we modeled the two tasks of code refinement and comment generation in a unified manner. We devise a novel architecture that comprises two interconnected models that synergistically collaborate through feedback to achieve both tasks.
By jointly training these models, our architecture may enable an efficient training process where the comment generation model gets feedback and guidance from the code refinement model. This has the potential to yield a significant improvement in state-of-the-art results for  comment generation.

\subsection{Code review automation using cross-task knowledge distillation}

Most deep learning models are trained using their own feedback by comparing their predictions to the true labels. Nevertheless, some approaches use feedback from another neural network for improving training. Like so, the training signal becomes more informative than a simple comparison between predictions and labels.
Jointly training two models can take the form of competition or collaboration.
Competitive models contest to outperform each other with respect to some metrics. Conversely, collaborative models cooperate, through a feedback mechanism, to achieve better performance. 

For instance, generative adversarial networks (GANs) are used for generative tasks wherein the goal is to create new data that resembles the original data distribution \cite{goodfellow2020generative}. The architecture of GANs typically comprises two deep learning networks: a generator and a discriminator. The former learns to create synthetic data, while the discriminator learns to distinguish between the generated and real data. GANs are trained in an adversarial and competitive manner where the generator tries to fool the discriminator while the latter tries to get better at differentiating between real and fake data. Consequently, the two models improve with training over time.

On the other hand, knowledge distillation is a collaborative learning technique that consists of transferring knowledge from a large model to a smaller one, as elucidated in \Sect{Sec:distillation}. Since we are addressing two different but dependent tasks in code review, we are particularly interested in cross-task knowledge distillation that permits the transfer of knowledge from one task to another.

Consequently, our main contribution is to use this technique (knowledge distillation) in the learning process of code reviews. Indeed, approaches like \cite{li2022codereviewer, tufano2022using, tufan2021towards} focus on training both models separately. Whereas, in our proposed architecture, we conjecture that producing code edits generation should be dependent on the reviews produced. Thus, we managed to provide feedback from the code edits generation model to the review generator one for improving the quality of the generated reviews. 

\begin{figure}[!htbp]
    \centering
    \includegraphics[width=1\linewidth]{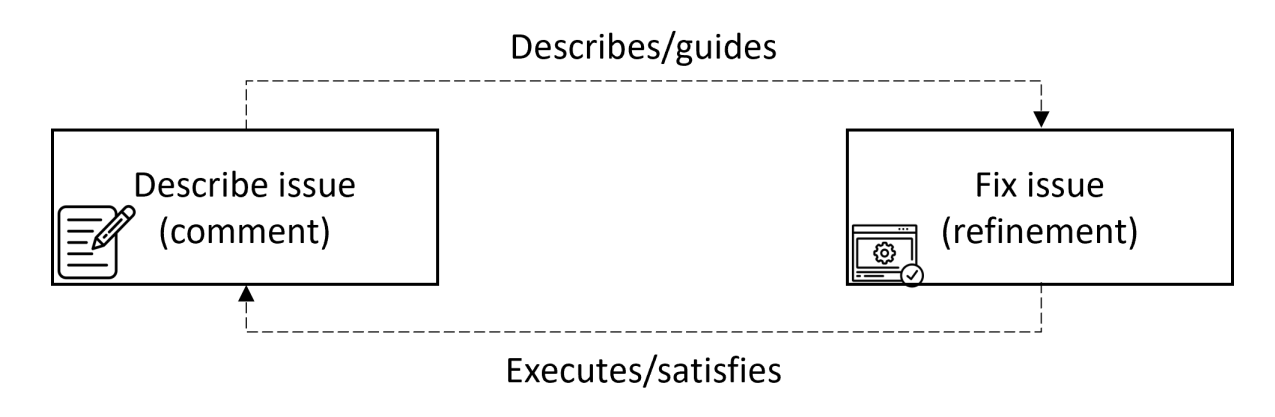}
    \caption{Relationship between the two tasks of code review, \ie comment generation and code refinement.}
    \label{fig:cr_relationship}
\end{figure}

In this contribution, we rely on the interdependence between the main two tasks of code review: comment generation and code refinement. We propose an architecture that trains two models simultaneously on these two tasks.
As shown in \Fig{fig:cr_relationship}, the \emph{comment generation task} and \emph{code refinement task} are tightly related.
The reviewer's comment guides the code refinement task and describes what the developer should do to improve/fix her code.
On the other hand, developers refine their code by executing the instructions of the review and trying to satisfy the reviewer's comments.
Despite their strong interdependence, these two tasks (\ie code review and code refinement ) have been addressed separately in the literature \cite{li2022codereviewer, tufano2022using, tufan2021towards}.
In our work, we rely on this relationship.
We use the code refinement task to guide, through feedback, the comment generation task and improve the quality of generations.

\Fig{fig:architecture} illustrates the architecture of our proposed model.
The first model (\ie student) generates a review (\ie comment) for the source code given as input.
The generated review, along with the code, is given as input to the second model (\ie teacher) to generate the necessary code edits that fix the input code with respect to the review. 
The loss of the teacher model backpropagates to the first model. That is, the teacher model gives feedback to the student model to indicate the relevancy of the generated review and to what extent it enabled the teacher model to generate the appropriate code edits.

We denote $\mathcal{D}$ our dataset that is composed of triplets $(c, r, c_r)$ where $c$ is the initial version of the source code submitted by the developer for review, $r$ is the review comment, and $c_r$ is a refined version of $c$ with respect to $r$.
$\mathcal{M}_t$ and $\mathcal{M}_s$ refer to the teacher and student models, respectively, such that: 

\begin{equation*}
\left\{
    \begin{array}{cc}
        \mathcal{M}_s:  & c \rightarrow r_p \\
        \mathcal{M}_t: & c, r_p \rightarrow c_{rp}
    \end{array}
\right.
\end{equation*}

The student model takes as input $c$ and generates the review $r_p$.
$r_p$ is fed along with $c$ to the teacher model that generates $c_{rp}$ trying to refine the source code.
The loss of the teacher model $\mathcal{L}_t$ is defined as follows:

\begin{equation*}
    \mathcal{L}_t = \mathcal{L}_{CE}(\mathcal{P}_{c_r}, \mathcal{P}_{c_{rp}})
\end{equation*}

It consists of comparing the predicted code edits to the real ones using the codeT5 loss function; \ie cross-entropy loss $\mathcal{L}_{CE}$ that compares the probability distribution $\mathcal{P}$ of the predicted output to the ground truth.
The formula that defines the cross-entropy loss for codeT5 is given by:
\begin{equation*}
    \mathcal{L}_{CE}(p, q) = -\sum_{i=1}^{N} p_i \log(q_i)
\end{equation*}
where:
\begin{itemize}
    \item $\mathcal{L}_{CE}$ represents the cross-entropy loss
    \item $N$ is the number of tokens in the sentence
    \item $y_i$ is the one-hot encoded vector representing the true $i^th$ token
    \item $q_i$ is the predicted probability vector of the $i^th$ token. The vector is of size $|vocabulary|$ and indicates for each word in the vocabulary its probability to be the $i^th$ token.
\end{itemize}

The loss of the student model $\mathcal{L}_s$ is defined as follows:

\begin{equation*}
    \mathcal{L}_s = \mathcal{L}_{CE}(\mathcal{P}_{r}, \mathcal{P}_{r_{p}}) + \mathcal{L}_t
\end{equation*}

It is a combination of the common cross-entropy loss of codeT5 model and the loss of the teacher model.
In short, the loss of $\mathcal{M}_s$ compares the predicted review to the real one, but also considers the feedback of the teacher.
That is, the student model will try not to only generate reviews that are close to the ground truth but also relevant reviews that enable the second model to predict the right code edits.

The total loss $\mathcal{L}_{1}$ of our proposed model \oapp~
 is defined as a combination of the two models' losses.

\begin{equation*}
    \mathcal{L}_1 = \alpha \mathcal{L}_s + \beta \mathcal{L}_t
\end{equation*}

$\alpha$ and $\beta$ indicate the importance of each loss.

\subsection{Embeddings alignment objective}

In natural language processing, hidden representations refer to intermediate representations of textual data learned by deep learning models during training.
These representations are not directly interpretable by humans, as they are typically high-dimensional vectors that capture complex patterns and relationships among words and phrases in the input text.

In the context of language models, each hidden layer may encode different aspects of the input text, such as syntactic structure, semantic meaning, or discourse coherence \cite{lopez2022ast}.
The encoding of each hidden layer can differ in several ways, such as the level of abstraction or granularity of the representations, the amount of contextual information captured, or the degree of specialization for a particular task or domain. For example, lower-level layers in a language model may encode more basic linguistic features, such as word-level embeddings or syntactic parse trees, while higher-level layers may capture more abstract semantic or discourse-level relationships between words and sentences.
Designing models that can leverage effective representations may have implications for downstream tasks such as text classification or language generation.

Usually, the last hidden layer of the encoder can be considered as an embedding for the input text that captures its meaning. This embedding is used by the decoder to generate the desired output.

In our approach, we are defining two models that generate reviews and code edits.
These two are closely related since the reviews are generally a description of the necessary code edits. 
Conversely, code edits represent typically the necessary changes to the code to satisfy the review comment.
Thus, a review comment and its corresponding code edits share similar meanings.
This tight relationship may suggest that there could be some shared underlying patterns or features, therefore similar representations, between the embeddings of reviews and code edits.

With the aim of improving the performance of our model, we add an \emph{embeddings alignment objective} where we try to align the embeddings of the review comment and the code edits.
The goal is to achieve close representations since the review comment and code edits share the same meaning.

\begin{figure*}[!htbp]
    \centering
    \includegraphics[scale=0.09]{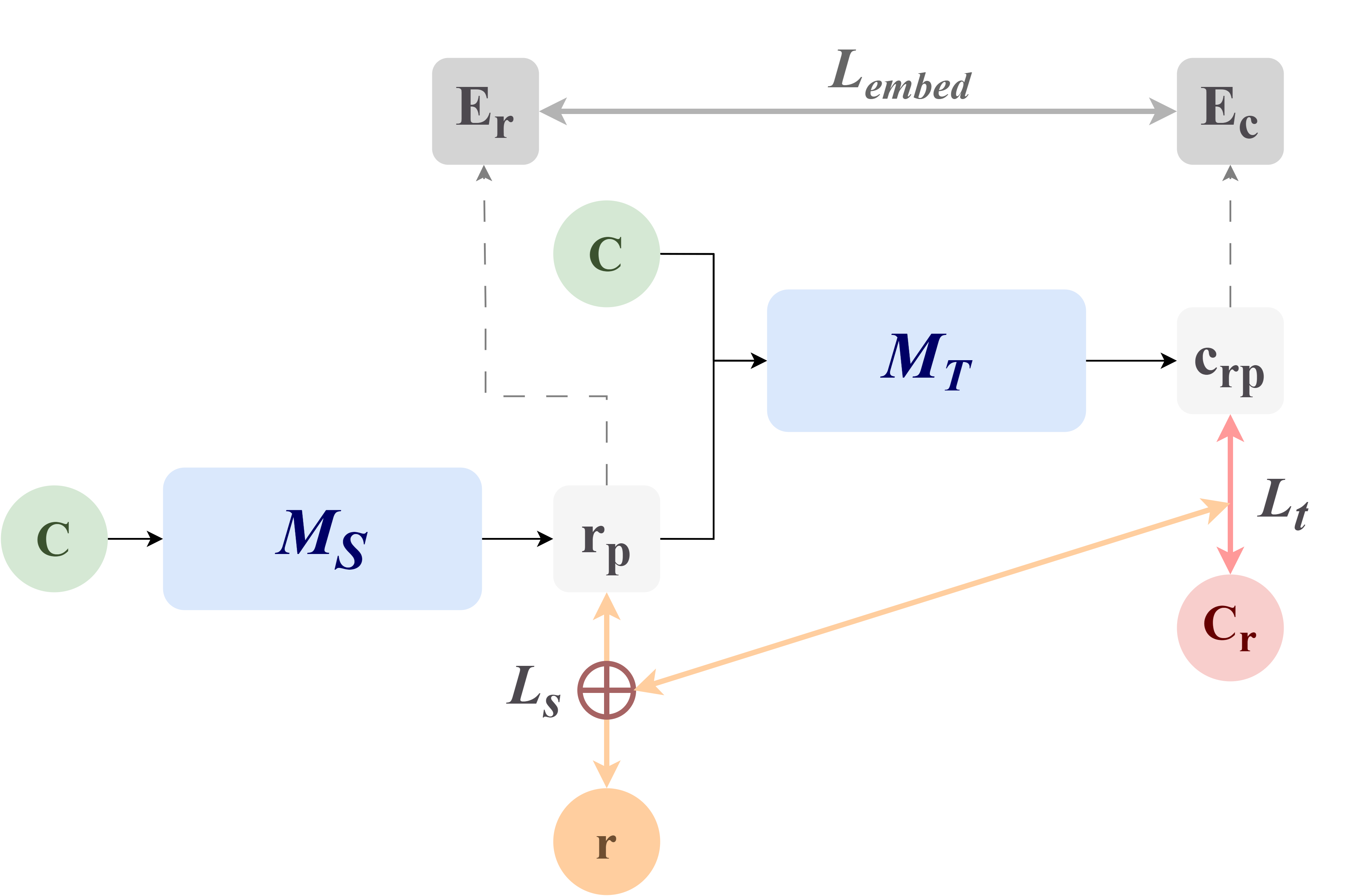}
    \caption{Detailed architecture of \oapp. \\
    The forward arrows represent the forward pass of the model, the circled shapes represent data (inputs, outputs, and ground truth), and the double arrows represent the losses.}
    \label{fig:architecture_details}
\end{figure*}

\Fig{fig:architecture_details} illustrates this objective.
We use codeT5 for both models $M_s$ and $M_t$, which is a transformer based on the encoder-decoder architecture. 
We use the representations generated by the encoder of $M_s$ as embeddings for the predicted review. Similarly, the embeddings generated by the encoder of $M_s$ are considered embeddings for the predicted revised version of the code.

Let us assume a predicted review $r_p$ and its corresponding predicted code edits $c_{rp}$ whose vector representations are $E_r$ and $E_c$ of dimension $n$. To align the embeddings, we define an objective $\mathcal{L}_{embed}$ to align the embeddings by minimizing the distance between the two vectors.
The distance between the two embeddings is defined as the mean square error (MSE) of the two embeddings:

\[ 
\mathcal{L}_{embed} = MSE(E_r, E_c) =  \frac{1}{n} \sum_{i=1}^n \left({E_c}_i - {E_r}_i\right)^2
\]

Our objective is to minimize the loss $\mathcal{L}_{embed}$, \ie the distance between the two embeddings, to achieve better and shared representations.
Thus, the total loss of \oapp~ is $\mathcal{L}_{2}$ defined as the combination of the student loss ($\mathcal{L}_{s}$), teacher loss ($\mathcal{L}_{t}$), and the embeddings alignment loss ($\mathcal{L}_{embed}$).

\[ 
\mathcal{L}_{2} = \alpha \mathcal{L}_{s} + \beta \mathcal{L}_{t} + \gamma \mathcal{L}_{embed}
\]

$\alpha$, $\beta$, and $\gamma$ are weights that indicate the importance of each loss.

\section{Evaluation}
\label{sec:evaluation}

We conducted a set of experiments to evaluate the performance of our proposed architecture. 
In this section, we present the addressed research questions followed by the experimental setting. Then, we present the results and discuss different threats related to our experiments.

\subsection{Research questions}

To evaluate our proposed approach, we define four main research questions:

\begin{itemize}
    \item \textbf{RQ1:} \textbf{Performance on comment generation}\\
    \emph{How does our model, with the simple combination of loss functions, perform compared to baseline models, on comment generation?}
    \item \textbf{RQ2:} \textbf{Performance on comment generation for each programming language}\\
    \emph{How does the same model perform, compared to baseline models, on comment generation with respect to each programming language?}
    \item \textbf{RQ3:} \textbf{Impact of the teacher pre-finetuning}\\
    \emph{To what extent does the pre-finetuning phase of the teacher contribute to enhancing the performance of the student model?}
    \item \textbf{RQ4:} \textbf{Effect of the embedding alignment objective}\\
    \emph{What is the impact of the embedding alignment objective on the performance of the model?}
\end{itemize}

\subsection{Experimental setting}
To address the different research questions, we conducted a set of experiments.

\paragraph{\textbf{Dataset}}
We use a dataset of $176 616$ code reviews obtained from \cite{li2022codereviewer}. 
The dataset has code review entries for nine ($9$) distinct programming languages.
\Table{tab:dataset} provides an overview of the dataset.
The data is split into $85\%$ for \emph{Train}, $7.5\%$ for \emph{Validation}, and $7.5\%$ for \emph{Test}. The data contains more than $1.3M$ lines of code.
To have a fair evaluation with \cite{li2022codereviewer} and \cite{tufano2022using}, we use the same sets of data. 
We cleaned the dataset and we kept the necessary features as explained below.
The resulting dataset is a set of triplets $(c, r, c_r)$ where $c$ is the initial version of the code, $r$ is the review comment, and $c_r$ is the revised source code after addressing the comment.

\newcommand{\bos}{<\!s\!>\!}
\newcommand{\eos}{\!<\!/s\!>\!}
\newcommand{\msg}{\!<\!msg\!>\!}
\newcommand{\pad}{\!<\!pad\!>\!}

\begin{table}
  \centering
  \caption{Distribution of the dataset over programming languages}
  \label{tab:dataset}
  \begin{tabular}{lccc}
    \toprule
        &   \textbf{Train} & \textbf{Validation} & \textbf{Test}\\
    \midrule
    \textbf{PHP}    & 7 979 & 973 & 1 032 \\
    \textbf{Ruby}    & 5 730 & 504 & 479 \\
    \textbf{C\#}    & 15 630 & 717 & 738 \\
    \textbf{C}    & 3 077 & 543 & 488 \\
    \textbf{Java}    & 31 288 & 2 177 & 2 206 \\
    \textbf{Python}    & 30 648 & 2 834 & 2 900 \\
    \textbf{C++}    & 13 274 & 1 362 & 1 308 \\
    \textbf{Go}    & 30 369 & 2 865 & 2 889 \\
    \textbf{JavaScript}    & 12 411 & 1 128 & 1 064 \\ \hline
    \textbf{Total}    & 150 409 & 13 103 & 13 104 \\
  \bottomrule
\end{tabular}
\end{table}

\paragraph{\textbf{Preprocessing}}
To properly train \oapp, the various inputs and outputs, for the different tasks, should be represented in a convenient way.

For the comment generation task, the student model $\mathcal{M}_s$ is fed with the code $c$ as input and generates a review $r_p$.
We clean the input code $c$ (\eg remove white spaces) and use RoBERTa tokenizer \cite{liu2019roberta} to split it into tokens.
\[
 c \Rightarrow c_1 c_2 \dots c_n
\]
The inputs are either truncated or padded with a special token $\pad$ to assert a fixed input length for $\mathcal{M}_s$ (the input length is a hyper-parameter that can be adjusted).
We add two other special tokens $\bos$ and $\eos$ to denote the beginning and the end of the input respectively.
The resulting input has this representation:
\[ 
c_1 c_2 \dots c_n \Rightarrow  \bos c_1 c_2 \dots c_n\pad\dots\pad\eos
\]

The output $r_p$ is also preprocessed in the same way; 
\[ 
r_p \Rightarrow \bos r_1 r_2 \dots r_p\pad..\pad\eos
\] 
where $r_i$ is the $i^{th}$ review token.

For the code refinement task, the teacher model $\mathcal{M}_t$ takes as input the predicted review along with the code and generates the revised version of the code.
The inputs are also cleaned, truncated or padded, and augmented with the special tokens.
The resulting input is represented as follows: 

\begin{flalign*}
&(c, r_p) \Rightarrow \bos c_1 c_2 \dots c_n\pad\dots\pad\msg 
r_{p_1} r_{p_2}\\ &\dots r_{p_n} 
\pad\dots\pad\eos
\end{flalign*}
where $\msg$ is a special token that separates the input code and review.
The output is preprocessed similarly as follows: 
\[ 
c_r \Rightarrow \bos c_{r_1} c_{r_2} \dots c_{r_p}\pad..\pad\eos
\]

\paragraph{\textbf{Experiments}}
To answer the defined research questions, we run different experiments.

For \emph{RQ1} and \emph{RQ2}, we fine-tune \oapp~on the training set using the losses $\mathcal{L}_t$ and $\mathcal{L}_s$. 
We run the fine-tuning in two phases: \emph{pre-finetuning} and \emph{finetuning} phases.
In the first phase, we fine-tune $\mathcal{M}_t$ on the \emph{code refinement} task (\ie predict the revised version of the code given the initial version).
This phase helps the teacher model acquire prior knowledge on this task, that is, providing the student model with good feedback.
In the second phase, the models $\mathcal{M}_s$ and $\mathcal{M}_t$ are fine-tuned jointly as illustrated by \Fig{fig:architecture}.
We assess our produced models on the test set and compare them to state-of-the-art works. 
To answer \emph{RQ2}, We evaluate also the performance of \oapp~with respect to each programming language.

In \emph{RQ3}, we investigate the impact of the pre-finetuning phase.
We run two experiments. First, we pre-finetune the teacher model $\mathcal{M}_t$ on the \emph{code refinement} task before the fine-tuning of \oapp.
Second, the \oapp~model is fine-tuned without pre-finetuning of $\mathcal{M}_t$.
We evaluate the produced models on the test set and compared the two alternatives. 

To address \emph{RQ4}, we implement the embedding alignment objective ($\mathcal{L}_{embed}$).
We run the pre-finetuning of the teacher. Then, we fine-tune \oapp~using $\mathcal{L}_t$, $\mathcal{L}_s$, and $\mathcal{L}_{embed}$.
$\mathcal{M}_t$ and $\mathcal{M}_s$ are codeT5 models that are based on the encoder-decoder architecture.
For each model, we extract the output of the last hidden layer of the decoder as embeddings.
$\mathcal{L}_{embed}$ is the $\emph{MSE}$ of the two embeddings.
The fine-tuned model is evaluated on the test set and compared to baseline models.

\paragraph{\textbf{Baseline models}}
To assess the effectiveness of our model, we compare \oapp~with state-of-the-art works: T5 \cite{tufano2022using}, CodeT5 \cite{wang2021codet5}, and CodeReviewer \cite{li2022codereviewer}.

In \cite{tufano2022using}, the authors fine-tuned T5 on generating comments from source code.
In \cite{li2022codereviewer}, the authors introduce \emph{CodeReviewer}, a codeT5 model that was pre-trained further on some tasks related to the code review, then fine-tuned on the comment generation task. The authors conducted a comparative evaluation of their results with \emph{CodeT5} \cite{wang2021codet5}, fine-tuned on the same task.

\paragraph{\textbf{Performance metrics}}
To assess the effectiveness of the models, we compute the Bilingual Evaluation Understudy (BLEU) score of the generated comments \cite{papineni2002bleu}.

The BLEU score is a widely used metric for evaluating the quality of text generated by deep learning models. 
It measures the degree of similarity between the generated text and a set of reference translations, in our case, the generated review and the actual review. A higher BLEU score indicates better text quality with respect to the references.
We use the BLEU-4 variant, that computes the n-gram overlap ($1\leq n\leq 4$) using this formula:
\[ 
BLEU\!-\!4 = min(1, \frac{output\_length}{reference\_length})(\prod_{i=1}^{4}precision_i)^{\frac{1}{4}}
\]
It computes the precision for n-grams of size 1 to 4 and adds a brief penalty for short sentences.

\paragraph{\textbf{Implementation}}
To implement the proposed models and run the different experiments, we used the PyTorch framework. 
We run the training on a machine  with four GPUs \emph{RTX 3090} and $24 GB$ of RAM per GPU. The models were trained using the Adam optimizer with a learning rate of $10^{-5}$, and we used cross-entropy as a loss function. 
The training process was performed on the training set for $30$ epochs, with a batch size of $16$. The validation set was used to monitor the model's performance and adjust the different hyperparameters accordingly.
Lastly, the test set was used in the final step to evaluate the performance of the models. 
For RQ1-3, we set the parameters $\alpha$ and $\beta$ to $0.5$.
For RQ4. we equally set the parameters $\alpha$, $\beta$, and $\gamma$ to $\frac{1}{3}$.


\subsection{Results}

\subsection*{Results for RQ1 - Performance on comment generation}
\Table{tab:results1} reports the results obtained from our first experiment to assess the performance of our model compared to state-of-the-art works.
In the pre-finetuning, the $M_t$ model achieved $81.79$, as the BLEU score, for the \emph{code refinement} task.
The produced model was used in the fine-tuning phase of \oapp~to guide, through feedback, the $M_s$ model on the \emph{comment generation} task.
As shown in \Table{tab:results1}, \oapp~outperforms the baseline models on both tasks.
As the difference is important for \emph{comment generation}, this suggests the effectiveness of our proposed feedback-based learning strategy where the pre-trained \emph{code refinement} model provides valuable guidance for the \emph{comment generation} task. 
By jointly optimizing both code review tasks, the \emph{comment generation} model is able to generate more accurate and relevant comments that not only align with the reviewers' feedback but also enable the \emph{code refinement} model to correctly fix the code.

While the difference is important for \emph{comment generation}, it is slight for \emph{code refinement}. 
This is attributed to the fact that the \emph{code refinement} model is trained on synthetic reviews that are generated by the \emph{comment generation} model. Since the generated reviews may contain grammatical errors and lack accuracy, the performance of the \emph{code refinement} model is affected, resulting in a minor improvement.
Conversely, our proposed feedback-based learning strategy enables the \emph{comment generation} model to benefit from the knowledge learned by the \emph{code refinement} model, resulting in an important improvement in generating accurate and relevant comments.

\begin{table}
  \centering
  \caption{Performance of \oapp~on comment generation compared to baseline models}
  \label{tab:results1}
  \begin{tabular}{l c c}
    \toprule
    \textbf{Model} & \textbf{Comment generation} & \textbf{Code refinement}\\
    \midrule
    \rowcolor{black!10} \multicolumn{3}{c}{\textcolor{black}{\textbf{Pre-finetuning}}}\\
        \oapp & & 81,79\\
    \rowcolor{black!10} \multicolumn{3}{c}{\textcolor{black}{\textbf{Fine-tuning}}}\\
        T5 & 4.39 & 77.03\\
        CodeT5 & 4.83 & 80.82\\
        CodeReviewer & 5.32 & 82.61\\
        \oapp & \textbf{6.68} & 82.84\\
    \bottomrule
\end{tabular}
\end{table}

\subsection*{Results for RQ2: Performance on comment generation for each
programming language}
\Table{tab:results2} shows the results of the performance of \oapp~model, compared to \emph{Code reviewer}, on the comment generation task for each programming language.

\begin{table*}
  \centering
  \caption{BLEU of \oapp, compared to baseline models, on comment generation per programming language}
  \label{tab:results2}
  \begin{tabular}{lccccccccc}
    \toprule
    \textbf{Language} & \textbf{PHP} & \textbf{Ruby} & \textbf{C\#} & \textbf{C} & \textbf{Java} & \textbf{Python} & \textbf{C++} & \textbf{Go} & \textbf{JavaScript}\\
    \midrule
    \textbf{CodeReviewer} & 8,1 & 5,49 & 5,92 & 5,16 & 4,03 & 4,36 & 5,71 & 6,05 & 5,24\\
    \bottomrule
    \textbf{\oapp} & \textbf{9.55} & \textbf{5.9} & \textbf{6.14} & \textbf{5.28} & \textbf{5.37} & \textbf{5.9} & \textbf{6.1} & \textbf{7.37} & \textbf{6.06}\\
\end{tabular}
\end{table*}

\oapp~outperforms \emph{CodeReviewer} on comment generation for all the programming languages.
However, this advantage is variable from one programming language to another.
This depends on the quality of the data for each programming language.
Also, software issues and best practices are different from one programming language to another.

We notice that \oapp~has a much better BLEU score for \emph{PHP}, and \emph{GO}.
However, the BLEU score is lower for \emph{C} and \emph{Java}.
This might depend on the usage and best practices for each programming language.
While some programming languages (\eg Java, C) have several best practices, coding standards, conventional issues, etc., developers are more tolerant and lenient with other languages (\eg PHP).
For instance, the Java community has a strong focus on code quality and best practices. There are many established coding standards and guidelines, such as the Java Code Conventions, which encourage developers to write high-quality, and maintainable code.
This could lead to more strict code review processes to ensure that these standards are upheld.

Moreover, the performance variability could be related to the nature and characteristics of the programming language.
Some programming languages are compiled, which means that the program will not be functional if it has errors.
which could lead to more stringent code review processes.
Other languages are interpreted, which means they are more tolerant as the program will still be functional and errors during run-time.

\subsection*{Results for RQ3: Impact of the teacher pre-finetuning}
In \emph{RQ3}, we investigate the impact of the teacher pre-finetuning phase on the performance of \oapp~on both code review tasks.
\Table{tab:results3} illustrates the influence of the pre-finetuning of the \emph{code refinement} model (\ie teacher) on the performance of \oapp.

For comment generation, there is an improvement of $0.46$ on the BLEU score when performing the pre-finetuning phase.
This demonstrates the importance of this phase as it enables the teacher model to acquire prior knowledge and have better performance on code refinement.
Thus, the teacher would provide the student with good and more relevant feedback during the fine-tuning phase. Consequently, the student model would be able to generate more accurate comments that align with the reviewers' comments.

Conversely, skipping the pre-finetuning phase produces less accurate results.
This is explained by the fact that the teacher model is less effective, so it provides the student with less accurate feedback.

Still, both experiments produce better results for comment generation compared to the literature.
This demonstrates the effectiveness of jointly addressing code review tasks using cross-task knowledge distillation.

These findings suggest that addressing several related tasks simultaneously using cross-task knowledge distillation might be beneficial even without having a very effective teacher model.
However, the pre-finetuning phase is important as it allows the teacher model to provide better and more informative feedback yielding better results than the student model.

\begin{table}
  \centering
  \caption{Impact of pre-finetuning the teacher (\ie code refinement model) on the performance of \oapp. }
  \label{tab:results3}
  \begin{tabular}{c c}
    \toprule
    \textbf{Comment generation} & \textbf{Code refinement}\\
    \midrule
    \rowcolor{black!10} \multicolumn{2}{c}{\textcolor{black}{\textbf{Without pre-finetuning}}}\\
        6,22 & 71.54\\
    \rowcolor{black!10} \multicolumn{2}{c}{\textcolor{black}{\textbf{With pre-finetuning}}}\\
        \textbf{6.68} & 82.84\\
    \bottomrule
\end{tabular}
\end{table}

\subsection*{Results for RQ4: Effect of the embedding alignment objective}
\Table{tab:results4} illustrates the impact of the embeddings alignment objective on the performance of our proposed model \oapp~compared to baseline models.
As shown in the table, this objective improved the performance of \oapp~for comment generation as measured by the BLEU score.
That is, since the revised code and the review share the same meaning, converging to close representations resulted in boosting the performance of \oapp~on code review.
The performance of \oapp~much improved on the \emph{comment generation} task.
However, we notice a slight decrease in the BLEU score of the \emph{code refinement} task.
This might be justified by the fact that the teacher is being trained on generated reviews that are not too accurate and might contain grammatical errors.
Additionally, only the teacher model provides feedback to the student, but it does not receive any feedback.
This explains the minor drop in the teacher's performance when tested on real reviews. Still, as our goal is to improve the comment generation, at the end of the process, we use the pre-finetuned model of the code refinement instead of one obtained after the joint fine-tuning.

These findings suggest that aligning the embeddings for different inputs that share similar semantics might be beneficial to the improvement of the model's performance on the downstream task.

\begin{table}
  \centering
  \caption{Influence of the \textit{embeddings alignment objective} on the performance of \oapp~for the comment generation task}
  \label{tab:results4}
  \begin{tabular}{l c c}
    \toprule
    \textbf{Model} & \textbf{Comment generation} & \textbf{Code refinement}\\
    \midrule
    \rowcolor{black!10} \multicolumn{3}{c}{\textcolor{black}{\textbf{Pre-finetuning}}}\\
        \oapp & & 81,79\\
    \rowcolor{black!10} \multicolumn{3}{c}{\textcolor{black}{\textbf{Fine-tuning}}}\\
        T5 & 4.39 & 77.03\\
        CodeT5 & 4,83 & 80,82\\
        CodeReviewer & 5,32 & 82,61\\
        \oapp & \textbf{7,33} & 80,96\\
    \bottomrule
\end{tabular}
\end{table}

\subsection{Threats to validity}
The evaluation results showed the effectiveness of our proposed architecture in addressing code review tasks. However, some threats might limit the validity of the presented evaluation results.

A first threat involves the nature of the data (\ie reviews) that are noisy and might contain non-English or misspelled words. This is mitigated by the use of codeT5, a state-of-the-art language model that uses Byte-Pair Encoding which is a subword-based tokenization algorithm. It breaks unseen words into several frequently seen sub-words that can be represented by the model.

A second threat entails the imbalance of the data, with some programming languages having fewer examples than others. 
This may lead to variable performance for each programming language. However, the use of a large pre-trained language model permits us to overcome this as CodeT5 is a large language model that is already trained on large amounts of code. Thus, fine-tuning this model on downstream tasks does not require a lot of data. Additionally, previous studies have demonstrated that the use of multilingual training datasets can result in superior model performance when compared to monolingual datasets for neural machine translation and code translation \cite{chiang2021breaking, zhu2022multilingual}, particularly for low-resource languages.

A last threat relates to the choice of hyperparameters that is crucial for the model's performance.
To have a fair evaluation with \cite{li2022codereviewer}, we just performed a grid search on the learning rate and batch size parameters.
For other hyperparameters, we use the same setting as in \cite{wang2021codet5}. However, it is likely that additional hyperparameter tuning would lead to further improvements.

\section{Related work}
\label{sec:literature}

Various techniques have been explored, in the literature, to support different tasks of code review, and these techniques can be classified into three categories: static analysis tools, similarity-based methods, and generative approaches.

In an effort to tackle the early stages of the code review process, a research direction has emerged that seeks to identify potential issues through the use of static analysis tools such as Checkstyle\cite{checkstyle}, PMD\cite{pmd}, and FindBugs\cite{hovemeyer2004finding, findBugs}. These tools, commonly referred to as linters, establish a set of rules that denote different types of issues related to security, code style, code quality, best practices, etc., and highlight parts of the code that are non-compliant with the defined rules.
Albeit the usefulness of such tools, the manual adaptation required to encompass most of the issues renders these tools rigid and less effective. Moreover, these tools need to be continuously calibrated with respect to issues and best practices that are variable over time and are affected by several factors such as software architecture, team composition, project characteristics, domain, and the like.
Hence, this inflexibility of static analysis tools diminishes their suitability for general adoption within software projects.

Other works employed similarity-based approaches to assist the code review process. These approaches assume that, in the context of code review, analogous situations may be encountered implying that similar issues may arise.
Consequently, historical knowledge can be exploited to hasten the code review process by resolving recurring issues.
Gupta et al. \cite{gupta2018intelligent} introduced DeepCodeReviewer (DCR), an LSTM-based model that is trained using positive and negative examples of (code, review) pairs. Given a new code snippet, a subset of candidate reviews is selected, from a predefined set of reviews, based on code similarity. Subsequently, DCR predicts a relevance score for each review with respect to the input code snippet and suggests reviews exhibiting high relevance scores.
Another work, proposed in \cite{siow2020core}, introduced a more sophisticated approach based on multi-level embedding to learn the relevancy between code and reviews. This approach uses word-level and character-level embeddings to achieve a better representation of the semantics provided by code and reviews.
These approaches have poor performance for unique code snippets and recommend irrelevant reviews in most cases. 

The last category, \ie generative approaches, attempted to use generative deep learning techniques to recommend reviews or code edits.
Tufano et al. \cite{tufan2021towards} proposed an approach that partially automates code corrections before and during code review. This approach is composed of two main components: a 1-encoder transformer encoder that recommends additional changes to the developer before submitting her code and a 2-encoder transformer that suggests the necessary code edits to satisfy the reviewer comments \cite{tufan2021towards}.
The authors improved their approach, in \cite{tufano2022using}, using T5, a pre-trained text-to-text transfer transformer \cite{raffel2019exploring}. To provide the model with prior knowledge of the downstream tasks, the authors pre-trained the T5 model on Java and technical English datasets.
Li et al. \cite{li2022codereviewer} pre-trained CodeT5 on four tasks, tailored specifically for the code review scenario, using a large-scale multilingual dataset of code reviews to better understand code differences and reviews. Then, the output model was fine-tuned on three downstream tasks: quality estimation (\ie accept/reject a pull request), review generation (\ie generate review comment), and code refinement (\ie recommend code edits to satisfy the reviewer comment).

Despite their tight relationship, these approaches addressed the different tasks of code review separately. In contrast, our approach employs code edits as feedback for the code review generation process. To achieve this, we rely on knowledge distillation introduced in \cite{hinton2015distilling}, where a model endowed with a greater number of parameters is employed to assist a smaller model in learning to produce decisions of comparable quality to the larger model, on the same task. More recently, cross-task distillation has been used in computer vision \cite{8890866, li2022prototypeguided, Ye_2020_CVPR}, allowing models to leverage knowledge gained from one task to improve performance on another related task. This is the latter that we use to improve, in our case, code review generation.


\section{Discussion and conclusion}
\label{sec:conclusion}
We propose a deep learning architecture based on cross-task knowledge distillation called \oapp.
Our architecture is composed of two models that are fine-tuned jointly on code review tasks: \emph{comment generation} and \emph{code refinement}.
The initial version of the code is given an input to the \emph{comment generation} model ($\mathcal{M}_s$) that predicts the review comment.
The predicted review is fed as input to the \emph{code refinement} ($\mathcal{M}_t$) model that generates the revised version of the code.
$\mathcal{M}_t$ provides feedback to $\mathcal{M}_t$ so that it generates reviews that are not only similar to the ground truth but also enable the \emph{code refinement} model to generate the appropriate code edits.
Additionally, we implemented an \emph{embeddings alignment objective} where we try to align the representations of the predicted review and the code edits since they share similar meanings.

We evaluated our model on a dataset of code reviews and we compared it to the state of the art.
Our results show that \oapp~outperforms baseline models.
Furthermore, we investigated the impact of the embeddings alignment objective on the performance of \oapp.
Our results demonstrate that this objective improves the model's performance further.
These findings suggest that simultaneously addressing several activities in some tasks might be beneficial as it allows to share of knowledge. This allows building collaborative models, that not only learn from the ground truth but also consider the feedback of other activities.
Our results for the \emph{embeddings alignment objective} demonstrate the importance of the inputs' representations to achieve good performance. Aligning the representations between inputs that share similar meanings might help to obtain better representations.

As part of future work, we aim to design another architecture with bidirectional feedback, in which both models collaborate to achieve better performance.
We plan to propose a generic framework that can be reused for other tasks with correlated activities. We intend also to integrate our model, as a bot, in some version control systems. Consequently, we plan to use this deployed model, in a user study, to assess its impact and usefulness in practice from the perspective of developers.

\bibliographystyle{IEEEtran}
\bibliography{references}

\end{document}